\input{aipcheck}

\documentclass[
    ,final            
  ]
  {aipproc}

\layoutstyle{8x11double}

\begin{document}

\title{Polyelectrolyte Adsorption on Charged Substrate}

\author{Chi-Ho Cheng}{
  address={Department of Physics and Center for Complex Systems,
National Central University, Taiwan, Republic of China.}
}

\author{Pik-Yin Lai}{
  address={Department of Physics and Center for Complex Systems,
National Central University, Taiwan, Republic of China.}
}


\begin{abstract}
The behavior of a polyelectrolyte adsorbed on a charged substrate of
high-dielectric constant is studied by both Monte-Carlo
simulation and analytical methods. It is found that in a low enough 
ionic strength medium, the adsorption transition is first-order where
the substrate surface charge still keeps repulsive. 
The monomer density at the adsorbed surface is identified as the
order parameter. It follows a linear relation with substrate
surface charge density because of the electrostatic boundary condition
at the charged surface. During the transition, the adsorption layer 
thickness remains finite. A new scaling law for the layer thickness
is derived and verified by simulation.

\end{abstract}

\maketitle


\section{Introduction}

The problem of polymer adsorption on an attractive surface has
drawn considerable interest due to its relation to surface effects
in critical phenomena and practical importance in technology and
biology. It is well established that the adsorption transition is
continuous if its attraction on the surface is short-ranged
\cite{degennes}. On the other hand,
long-ranged electrostatic interactions in polyelectrolyte systems
pose many challenging theoretical problems.
Recently the macroion adsorption on an electrostatically
attractive interface \cite{gelbert,review2} and the
associated charge inversion phenomena of adsorbed polyelectrolytes
\cite{joanny,nguyen} acquire lots of attention.

Previous analytical approaches impose the continuity of monomer 
density across the charged surface, the surface monomer density is 
then set to zero \cite{wiegel,muth}. The polyelectrolyte
adsorption problem becomes the competition between the electrostatic
force from charged surface and the configurational entropy
of the polyelectrolyte itself. The transition has been shown to
be continuous. However, we found that the above treatment is not
quite adequate since the electrostatic boundary condition at the
charged surface is not faithfully respected. Furthermore, the 
electrostatic boundary condition may induce different physics
in the presence of a substrate under the charged surface.
In this paper, we study the adsorption of a polyelectrolyte 
on a high-dielectric substrate at low ionic strength
(e.g. aqueous solution with a metal substrate) 
in which image charge interaction is attractive. 
The adsorption transition occurs where the
surface charges are repulsive instead of the attractive case that
were usually studied. The problem is tackled by 
Monte-Carlo simulations and analytical methods 
taking full account of appropriate boundary
conditions. It is found that the order of adsorption
transition, the physical mechanism, and the scaling behavior are
all different from those of the attractive surfaces.

\section{Model}

A polyelectrolyte carrying positive charges is immersed in a
medium ($z>0$) of dielectric constant $\epsilon$. At $z=0$ there
is an impenetrable surface of uniformly surface charge density
$\sigma$. Below that ($z<0$), it is a substrate of
dielectric constant $\epsilon'$. Denote the charge on a polymer
segment $ds$ by $q_0 ds$, the Hamiltonian is written as
\begin{eqnarray}
{\cal H} &=&  \frac{3k_{\rm B}T}{2l_0^2} \int_0^N ds
\left(\frac{\partial{\vec r(s)}}{\partial{s}}\right)^2 
+\frac{1}{2} \int_0^N ds \int_0^N ds' \{
\nonumber \\
&&
\Gamma\frac{{\rm
e}^{-\kappa|\vec r(s) - \vec r(s')|}}{|\vec r(s) - \vec r(s')|} 
- \Gamma'(2-\delta_{s,s'})\frac{{\rm e}^{-\kappa|\vec r(s) - \vec
r'(s')|}}{|\vec r(s) - \vec
r'(s')|} \} \nonumber \\
&&+ h \int_0^N ds \kappa^{-1}{\rm e}^{-\kappa \vec r(s)\cdot\hat
z} 
\nonumber \\
&&+ \omega\int_0^N ds\int_0^Nds' \delta(\vec r(s)-\vec r(s'))
\label{ham}
\end{eqnarray}
where $s$ is the variable to parametrize the chain, $l_0$ the bare
persistence length, and $\kappa^{-1}$ the Debye screening length.
$\vec r(s)=(x(s),y(s),z(s))$, $\vec r'(s')=(x(s'),y(s'),-z(s'))$
are the positions of the monomers and their electrostatic images,
respectively.  $\Gamma = q_0^2/\epsilon$, $\Gamma' =
\Gamma(\epsilon'-\epsilon)/(\epsilon'+\epsilon)$, and $h=4\pi
q_0\sigma/(\epsilon'+\epsilon)$ are the coupling parameters
governing the strengths of Coulomb interactions among the monomers
themselves, between the polymer and its image, and between the
polymer and the charged surface, respectively. In good solvent regime,
$\omega>0$ in the last term.

\section{Monte-Carlo Results}

The continuum model is discretized to perform Monte-Carlo
simulations. $\vec r(s)$ is replaced by a chain of beads $\vec
r_i$ ($i=1,\ldots,N$) with hard-core excluded volume of finite
radius $a$.  
Polymer lengths up to $N=120$ are employed. The units
of length and energy are $2a$ and $q_0^2/2\epsilon a$,
respectively. The ratios $\epsilon'/\epsilon$ are studied from 2
to 12.5 (aqueous solution with a metallic substrate). Runs up to
$10^9$ MC steps are performed.

\begin{figure}
\includegraphics[height=.25\textheight]{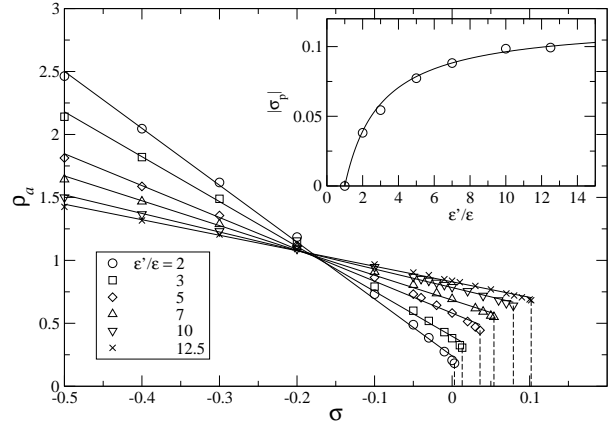}
\caption{Monte-Carlo results for the normalized monomer density 
at the surface,
$\rho_a$, as a function of surface charge density $\sigma$ (in
unit of $q_0/4a^2$) for different $\epsilon'/\epsilon$ at
$\kappa^{-1}=25$. The fitted straight lines are terminated at
their adsorption transition points. The vertical dashed lines are
drawn as guides to the eyes. Inset: The polarization surface
charge density due to the charged polymer, $|\sigma_{\rm p}|$, as
a function of $\epsilon'/\epsilon$. $\sigma_{\rm p}$ saturates as
$\epsilon'/\epsilon\rightarrow\infty$. The sign of $\sigma_{\rm
p}$ is opposite to $q_0$ and is thus negative. The solid curve is
fitted from Eq.(\ref{sigmapoly}) with $\sigma_{\rm poly}=0.118$. }
\label{adsorb6.eps}
\end{figure}

The adsorption behavior can be characterized by normalized
monomer density $\rho(z)$. $\rho_a\equiv\rho(z=a)$, the probability
density of monomers adsorbed on the substrate, is chosen as the order 
parameter to describe the adsorption transition. 
$\rho_a$  as a function of the surface charge density $\sigma$ for
various $\epsilon'/\epsilon>1$ is shown in Fig.\ref{adsorb6.eps}.
It is seen that $\rho_a$ vanishes abruptly when $\sigma$ 
increases up to its threshold value $\sigma_{\rm t}$ indicating a
first-order transition. At low enough ionic strength that $\kappa^{-1}$ 
is much larger than absorption layer thickness, $\sigma_{\rm t}>0$
for $\epsilon'/\epsilon > 1$.
The discontinuous drop of $\rho_a$ across the transition decreases to
zero as $\epsilon'/\epsilon\to 1$.
We have also verified that the sharp jump in energy (latent
heat) across the transition is proportional to $N$ as
expected for a first-order transition. Same results are obtained
for larger $\kappa^{-1}$.

Furthermore, $\rho_a$ is linear in $\sigma$ with the slope 
depending on $\epsilon'/\epsilon$. 
Such a linear relation between $\rho_a$ and $\sigma$ can be understood 
from the electrostatic boundary conditions that the system has to satisfy. 
The electric potential $\phi(z)$ in the neighborhood of $z=0$ boundary
obeys
\begin{equation}
-\left.\frac{\partial \phi}{\partial z} \right|_{z=0^+} +
 \left.\frac{\partial \phi}{\partial z}\right|_{z=0^-}
 = -\frac{4\pi}{\epsilon}\left(
\frac{2\sigma}{\epsilon'/\epsilon+1}+\sigma_{\rm p}\right)
 \label{boundary1}
\end{equation}
where $\sigma_{\rm p}$ is the polarization surface charge density
induced by the polyelectrolyte only. Notice that $\sigma_{\rm p}$
depends only on $\epsilon'/\epsilon$. It is independent of $\sigma$ 
in the adsorbed regime near the transition. The reason is based on
the electric blob picture which will be explained in next section.
If one treats the polyelectrolyte as a marcomolecule with a
well-defined surface, its surface charge density at $z=a$ should
be proportional to the monomer density $\rho_a$. It also applies
to the electric field in the $z<0$ region. One have
\begin{eqnarray}
K \left.\frac{\partial \phi}{\partial z}\right|_{z=a^-} &=&
-\frac{4\pi}{\epsilon}\rho_a , \\ K \left.\frac{\partial \phi}
{\partial z}\right|_{z=0^-}  &=&
-\frac{4\pi}{\epsilon}\frac{2\epsilon'}{\epsilon'+\epsilon}\rho_a
 \label{boundary2}
\end{eqnarray}
where $K>0$ is the proportionality constant. Applying
the electric field continuity from $z=0^+$ to
$z=a^-$, and using Eqs.(\ref{boundary1})-(\ref{boundary2}), one
get the linear behavior
\begin{equation}
\rho_a = -\frac{2K}{\epsilon'/\epsilon-1} \left(\sigma +
\frac{\epsilon'/\epsilon+1}{2}\sigma_{\rm p} \right)
\label{rho-sigma}.
\end{equation}
Notice that $K$ and $\sigma_{\rm p}$ depends only on
$\epsilon'/\epsilon$. As shown in Fig.\ref{adsorb6.eps}, 
the slope $K$ decreases monotonically with $\epsilon'/\epsilon$. 
Lines with different values of $\epsilon'/\epsilon$ 
intersect at a common point implies the naive intuition that 
a more polarizable substrate gives a stronger adsorption 
is not always true.

Without substrate surface charge, set $\sigma=0$ in Eq.(\ref{rho-sigma}), 
one get
\begin{equation}
\sigma_{\rm p}= -\frac{\rho_a|_{\sigma=0}}{K}
\frac{\epsilon'/\epsilon-1}{\epsilon'/\epsilon+1}. \label{sigmapoly}
\end{equation}
The dependence on $\epsilon'/\epsilon$ can be considered as a flat
surface of surface charge density 
$\sigma_{\rm poly}\equiv \rho_a|_{\sigma=0}/K$ located above the
substrate. $K$ are obtained by linear fittings in Fig.\ref{adsorb6.eps},
and hence $\sigma_{\rm p}$ as a function of $\epsilon'/\epsilon$ is
plotted in the inset. $\sigma_{\rm p}$ increases from zero to a saturated
value as $\epsilon'/\epsilon\rightarrow\infty$.
Presumably, Eq.(\ref{sigmapoly}) suggests that it is possible to
experimentally detect the sign and magnitude of the charge of the
polyelectrolyte by measuring the substrate polarization.

\section{Analytical Results}

At low ionic strength, a polyelectrolyte can be treated as 
electric blobs arranged longitudinally. The rod-like polyelectrolyte 
tends to lie down on the substrate so as to lower its energy. 
It is thus rigid (rod-like) in x-y plane, but
still flexible in the z-direction.  In the
adsorbed regime, decreasing $\sigma$ (chain is attracted more by
the surface) would cause the rearrangement of the electric blobs
such that the fluctuations in the z-direction is reduced (and
hence the layer thickness decreases), but the fluctuations in the
other two directions and the size of the blobs are basically
unchanged. The effective in-plane surface charge distribution of
the polyelectrolyte, and hence $\sigma_{\rm p}$, is not affected 
by $\sigma$. The excluded volume effect is ignored because it takes 
almost no effect in z-direction. The effect from self-electrostatic 
interaction can be absorbed by renormalising the bare persistence 
length $l_0$ to $l$.

Because the monomer would feel the strongest attraction from its
direct image around the adsorption regime, the $\Gamma'$-term in
Eq.(\ref{ham}) is approximated by the interaction of every monomer
and its corresponding image only. The residual attraction from
the images of other monomers could be absorbed into the coupling
parameter $\Gamma'$ by renormalising $q_0$ to $q$. The partition
function is reduced to
\begin{eqnarray} \label{partit}
Z &=& \int{\cal D}[\vec r(s)]\exp  [\int_0^N ds \{
-\frac{3}{2l^2}
 \left(\frac{\partial{\vec r(s)}}{\partial{s}}\right)^2 
\nonumber \\
&& + \frac{\beta\Gamma'}{4}
 \frac{{\rm e}^{-2\kappa \vec r(s)\cdot\hat z}}{\vec r(s)\cdot\hat z}
 - \beta h \kappa^{-1} {\rm e}^{-\kappa \vec r(s)\cdot\hat z}
 \}].
\end{eqnarray}
$\vec r(s)$ is transformed to 
$\rho(\vec r)\equiv\frac{1}{N}\int_0^Nds \delta(\vec r-\vec r(s))$
by introducing an auxiliary field. Ground state dominance 
in large-$N$ limit is then applied.
By variational principle, one obtains the
Edwards-Schr\"{o}dinger equation,
\begin{equation} \label{sch}
\left(-\frac{l^2}{6}\frac{d^2}{dz^2} -\frac{\beta\Gamma'}{4}
\frac{{\rm e}^{-2\kappa z}}{z} + \beta h \kappa^{-1} {\rm
e}^{-\kappa z} \right) \psi(z) = \varepsilon_0 \psi(z)
\label{tise}
\end{equation}
where $\varepsilon_0$ acts as a Lagrange multiplier to enforce the
constraint of the ground state wavefunction normalization. The
monomer density is given by $\rho(z)=|\psi(z)|^2$. Eq.(\ref{tise})
also describes a quantum particle at its ground state moving under
a combined potential of a 1d screened Coulomb attraction and an
almost linear potential. However, the boundary condition expressed
by Eq.(\ref{rho-sigma}) is different from that of the hard-wall one
$\psi|_{\rm s}=0$ usually employed for the quantum particle.
Instead $\psi|_{\rm s} = \sqrt{\rho_a} \neq 0$ for the present problem
implies that the steric force felt by the polyelectrolyte from the
charged surface should be modified \cite{steric}. 
Setting $\psi|_{\rm s}=0$ \cite{wiegel,muth,review1} 
in previous studies is not completely correct.

\begin{figure}
\includegraphics[height=.25\textheight]{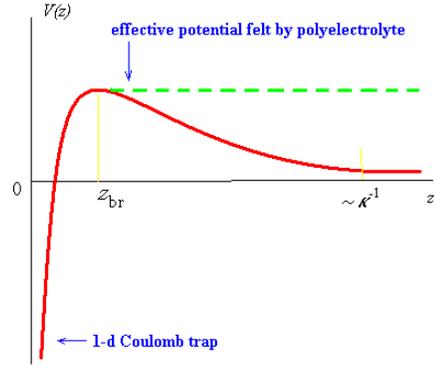}
\caption{Effective Potential felt by the polyelectrolyte}
\label{poly04a.eps}
\end{figure}

During the adsorption, the rod-like polyelectrolyte tends to lie
down on the charged surface. The thickness of the adsorption layer
is of the same order of the gyration radius in z-direction. At low
ionic strength corresponding to Debye length much larger than 
layer thickness, the polyelectrolyte can only feel the potential
barrier formed by $V(z)$ as shown in Fig\ref{poly04a.eps}. 
Analytically, the original potential $V(z)$ 
in Eq.(\ref{sch}) is replaced by
\begin{equation} \label{potential}
V_{{\rm mod}}(z) = \cases{  +\infty , &$\qquad z < a $ \cr
 V(z)
, &$\qquad a \leq z < z_{\rm br}   $\cr
 V(z_{\rm br}), &$\qquad  z \geq z_{\rm br}
$\cr }
\end{equation}
where $z_{\rm br}$ is chosen such that $V'(z_{\rm br})=0$, and
$V(z_{\rm br})$ is the barrier height. In the limit of $\sigma =
\kappa=0$, the analytic solution exists in the form
\begin{equation}
\psi(z) = {\rm
W}_{\lambda,1/2}(\frac{3\beta\Gamma'}{2l^2\lambda}z)
 \hspace{1mm};\hspace{1mm}
\varepsilon_0 =
-\frac{3\beta^2\Gamma'^2}{32l^2}\frac{1}{\lambda^2}
\label{binderg}
\end{equation}
where ${\rm W}_{\lambda,1/2}$ is the Whittaker's notation of the
confluent hypergeometric function \cite{1dcoulomb}, and $\lambda$
is the least value satisfying the boundary condition. Bound state
exists for arbitrary $\epsilon'/\epsilon > 1$. It implies
$\sigma_{\rm t} > 0$ at low enough ionic strength.

\begin{figure}
\includegraphics[height=.24\textheight]{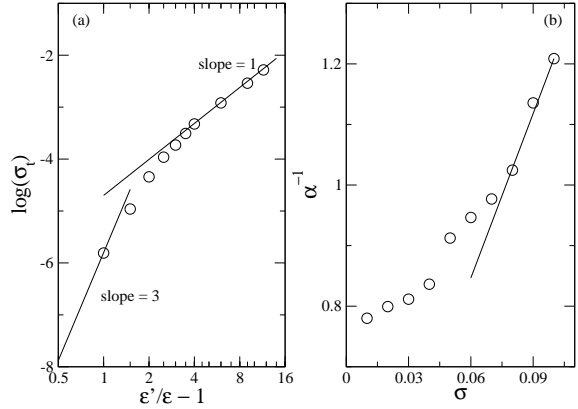}
\caption{
(a) Simulation results for the surface charge density at the
transition, $\sigma_{\rm t}$, as a function of
$\epsilon'/\epsilon$ in logarithmic scale at $\kappa^{-1}=25$. The
straight lines indicate slopes of 1 and 3 as suggested in the text.
(b) Simulation results for the decay length $\alpha^{-1}$
(which is proportional to the layer thickness) as a
function of $\sigma$ (in unit of $q_0/4a^2$) for
$\epsilon'/\epsilon=12.5$, $\kappa^{-1}=25$. The straight line is
a linear fit to points near the transition. 
$\sigma_{\rm t}=0.102$ in this case.  $\alpha^{-1}$ is
obtained from exponential fitting to the tail of corresponding
density profile.}
\label{sigma3.eps}
\end{figure}

No exact solution exists for $\sigma, \kappa > 0$ in
general but one can analyze it around the transition. In the
region of interest, the Coulomb term dominates
over the almost linear term. One can approximate the
binding energy by Eq.(\ref{binderg}). The
polyelectrolyte undergoes a de-sorption transition when the
binding energy is equal to barrier height $V(z_{\rm br})$.
After some algebra, it shows $\sigma_{\rm
t}\sim(\epsilon'/\epsilon-1)$ for $\epsilon'/\epsilon\gg 1$ and
$\sigma_{\rm t}\sim(\epsilon'/\epsilon-1)^3 $ for
$\epsilon'/\epsilon \sim 1$. This analytic result is consistent
with simulation as shown in Fig.\ref{sigma3.eps}a.

An approximate solution for density profile $\rho(z)$ for 
$\sigma>0$ can be studied by the variational wavefunction
\begin{equation}
\psi(z) = \sqrt{\rho_a}(1+\mu\alpha(z-a))
 {\rm e}^{-\frac{1}{2}\alpha(z-a)}
\end{equation}
where $\alpha^{-1}$ is the decay length. $\mu$ is positive because
the wavefunction is restricted to be nodeless. $\alpha$ and
$\mu$ are not independent but related via the wavefunction
normalization condition. The inverse decay length is calculated
to be
\begin{equation} \label{decay}
\alpha= \frac{3\beta\Gamma'}{2l^2} + \rho_a
\end{equation}
where the leading term is independent of $\sigma$. Near the
transition, the decay length $\alpha^{-1}$ and hence 
the adsorption layer thickness $D\sim\alpha^{-1}$ remains
finite. 
Combining Eqs.(\ref{rho-sigma}) and (\ref{decay}), one get the scaling
behavior
\begin{equation} \label{Ds}
D_{\rm t}-D 
\sim (\alpha_{\rm t}^{-1}-\alpha^{-1})
\sim (\sigma_{\rm t}-\sigma) 
\end{equation} 
where $D_{\rm t}$ is the threshold layer thickness.
$\alpha^{-1}$ as a function of $\sigma$ from simulations 
is shown in Fig.\ref{sigma3.eps}b.
 
It would be beneficial to compare with the case that
a polyelectrolyte is adsorbed onto an attractive charged
substrate of low-dielectric constant (e.g. DNA in aqueous solution 
onto a charged lipid membrane).
Its asymptotic solution for $z\to+\infty$ to Eq.(\ref{sch})
reproduces the usual scaling 
$D\sim\alpha^{-1}\sim|\sigma|^{-\frac{1}{3}}$ \cite{joanny}.
The adsorption onto a low-dielectric substrate is continuous.
The order parameter $\rho_a$ vanishes continuously across 
the transition, and the layer thickness swells to infinity as the 
polyelectrolyte is de-sorbed \cite{lowdielectric}.

\section{Discussion}


A strongly charged polyelectrolyte immersed in a salt solution
will attract oppositely charged ions to condense until its
effective charge density reaches the Manning threshold
\cite{manning}. This means that one can just renormalize $q_0$ in
our system to $2ea/l_{\rm B}$ if $q_0$ is larger than $2ea/l_{\rm
B}$ ($l_B$ is the Bjerrum length). Similarly, the strongly charged
surface of bare charge density larger than $\kappa/(\pi l_{\rm
B})$ is just renormalized back to $\kappa/(\pi l_{\rm B})$
\cite{bocquet}. 


Although we focus on the adsorption of a single
polyelectrolyte, our results may provide a starting point to
study the charge inversion and multi-layer adsorption.
\cite{dobrynin}.
At low ionic strength, polyelectrolytes are
adsorbed in a multi-layer structure because of strong Coulomb
repulsion. Each layer is composed of the parallel 1d Wigner
crystal \cite{shklovskii}. From our physical picture of a single
polyelectrolyte adsorption, the upper bound for the thickness of
the multi-layer structure is 
$z_{\rm br}\sim\sigma^{-1/2}(\epsilon'/\epsilon-1)^{1/2}$.  
This suggests one can easily control just one single layer 
adsorbed onto high-dielectric substrate by tuning 
the surface charge density.
Rigorous treatment based on this physical picture will be
elaborated elsewhere.




\bibliographystyle{aipproc}   


\bibliography{sample}

\IfFileExists{\jobname.bbl}{}
 {\typeout{}
  \typeout{******************************************}
  \typeout{** Please run "bibtex \jobname" to optain}
  \typeout{** the bibliography and then re-run LaTeX}
  \typeout{** twice to fix the references!}
  \typeout{******************************************}
  \typeout{}
 }

\end{document}